\documentclass[twocolumn,aps,showpacs,superscriptaddress]{revtex4}

\usepackage{graphicx}

\newcommand{\BEQ}{\begin{equation}}
\newcommand{\EEQ}{\end{equation}}
\newcommand{\BEA}{\begin{eqnarray}}
\newcommand{\EEA}{\end{eqnarray}}
 
\newcommand{\p}{\partial}

\begin{document}
\title{Thermodynamic first order transition and inverse freezing
in a 3D spin-glass.}

\date{\today} 
\author{M. Paoluzzi} 
\affiliation{IPCF-CNR, UOS Roma, 
  P.le Aldo Moro 2, I-00185 Roma, Italy}
\affiliation{Dipartimento di Fisica,  Universit\`a di Roma 3,
Via della Vasca Navale 84, I-00146 Roma, Italy}

\author{L. Leuzzi} 
\affiliation{IPCF-CNR, UOS Roma,  P.le Aldo Moro 2, I-00185 Roma, Italy}
\affiliation{Dipartimento di Fisica,  Universit\`a "Sapienza",
  P.le Aldo Moro 2, I-00185 Roma, Italy}

\author{A. Crisanti}
\affiliation{Dipartimento di Fisica, Universit\`a "Sapienza",
  P.le Aldo Moro 2, I-00185 Roma, Italy}


\begin{abstract}
We present a numerical study of the random Blume-Capel
model in three dimension. The phase diagram is characterized by
spin-glass/paramagnet phase transitions both of first and second order
in the thermodynamic sense. Numerical simulations are performed using
the Exchange-Monte Carlo algorithm, providing clear evidence for
inverse freezing.  The main features at criticality and in the phase
coexistence region are investigated.  We are not privy to other
 3D short-range systems with
quenched disorder undergoing inverse freezing.
\end{abstract}

\maketitle

\begin{figure}[t!] 
\centering
\includegraphics[width=.9\columnwidth]{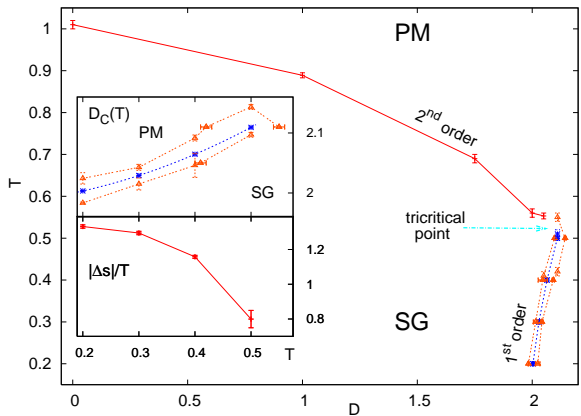}
\caption{Phase diagram in $D,T$: second order transition and an
inverted FOTP occur. In the latter case also the spinodal lines are
reported (dashed). Bottom inset: latent heat $|\Delta s|/T$ along the
first order line.  Top inset: detail of IF region, interpolation of
transition line $D_c(\infty,T)$ (dotted), spinodal lines (dashed). The
error bars are the FSS of the minimal interval in $T$ and $D$ at each
$L$ needed to identify the crossings in $\xi_c/L$ curves (for continuous
transitions) or compare the areas under $P_{N}(\rho)$ for FOPT.}
\label{fig:phdi}
\end{figure}
\begin{figure}[t!] 
\centering
\includegraphics[width=.99\columnwidth]{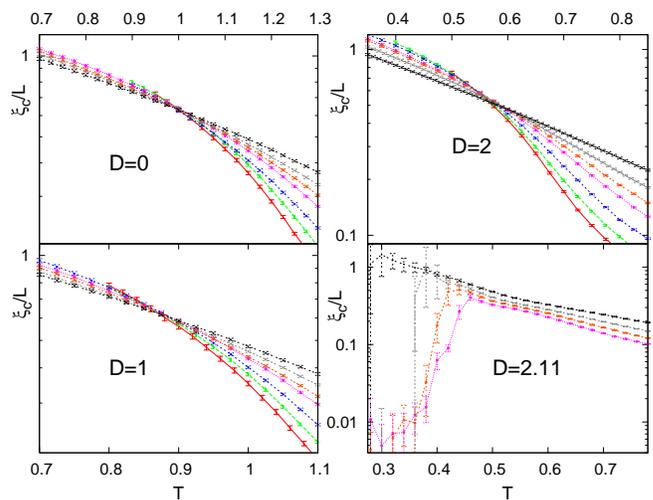}
\caption{Scaling funtions $\xi_c/L$ vs. $T$ for different values of
the chemical potential $D$. For $D=0,1,2$ ($L=6,8,10,12,16,20,24$) a
continuous phase transition is found in the region of scale
invariance. At $D=2.11$ ($L=6,8,10,12$) no crossing is observed and at
low $T$ $\xi_c/L\to 0$. 
}
\label{fig:xisuL}
\end{figure}

 {\em Introduction.} We aim to investigate the phenomenon of
reversible inverse transition (IT), occurring between a solid and a
liquid in the inverse order relation relatively to standard
transitions. The case of ``ordering in disorder'', occurring in a
crystal solid that liquefies on cooling, is generally termed {\em
inverse melting}.  If the solid is amorphous the IT is termed {\em
inverse freezing} (IF).
\\ \indent IT's are observed in different materials.  The first
examples were the low temperature liquid and crystal phases of helium
isotopes He$^3$ and He$^4$ \cite{Wilks87}.  A more recent and complex
material is methyl-cellulose solution in water, undergoing a
reversible inverse sol-gel transition \cite{Chevillard97}. Other
examples are found in poly(4-methylpentene-1) - P4MP1 - at high
pressure \cite{RHKMM99}, in solutions of $\alpha$-cyclodextrine
($\alpha$CD) and 4-methypyridine (4MP) in water \cite{Plazanet}, in
ferromagnetic systems of gold nanoparticles \cite{Donnio07} and for
the magnetic flux lines in a high temperature superconductor
\cite{Avraham01}.
In mentioning these cases, we stick to a definition of IT as the one
hypothesized by Tammann \cite{Tammann} a century ago: a reversible
transition in temperature at fixed pressure \cite{footnote1} whose low
$T$ phase is an isotropic fluid (or paramagnet - PM - for magnetic
systems).
IT is not an exact synonim of reentrance.  Indeed,
a reentrance can be absent, as for $\alpha$CD
\cite{Plazanet} or methyl-cellulose
\cite{Chevillard97} solutions, where no high temperature fluid
phase is detected. Moreover, not all reentrances are signatures
of an IT, as for those phases with different kind of symmetry separated by
reentrant isobaric  transition lines in temperature, cf.,
e.g., Refs.
\cite{Cladis75}, in which, however, no
melting to a completely disordered isotropic phase is present.
\\ \indent A thourough explanation of the fundamental mechanisms
leading to the IT would require a microscopic analysis of the single
components behavior and their mutual interactions as temperature
changes accross the critical point.  Due to the complexity of the
structure of polymers and macromolecules acting in such
transformations a clear-cut picture of the state of single components
is seldom available. For the case of methyl-cellulose
\cite{Chevillard97}, where methyl groups (MGs) are distributed randomly
and heterogeneously along the polymer chain \cite{Arisz95}, Haque and
Morris \cite{Haque93} proposed that chains exist in solution as folded
hydrophilic bundles in which hydrophobic MGs are packed.  As $T$ is
raised, bundles unfold, exposing MGs to water molecules and causing a
large increase in volume and the formation of hydrophobic links
eventually leading to a gel condensation. The polymers in the folded
state are poorly interacting but also yield a smaller entropic
contribution than the unfolded ones. A similar behavior has been recently modeled in colloidal systems by the so-called Janus particles \cite{Sciortino09}.
\\ \indent
Under the assumption that this is at least one of the fundamental mechanisms underlying IT, we model it approximating the folded/unfolded conformation by
 bosonic spins: $s=0$ representing inactive state, $s\neq 0$
interacting ones, cf. Refs. \cite{SSPRL04,SS05}. To represent
the randomness on the position of the "interaction carrying" elements
(e.g., MGs) we will introduce quenched disorder.
 Theoretical modeling for IT's mainly consists in heuristic
reproductions of the phenomenon \cite{Feeney,SSPRL04}. In
particular, IF, has been recently observed in spin-glass
(SG) mean-field (MF) Blume-Emery-Griffiths-Capel (BEGC) models with
spin$-1$ variables \cite{CLPRL05, Sellitto06}.
We will focus on the random Blume-Capel (BC) model \cite{GSJPC77},
whose MF solution \cite{CLPRL02} predicts a phase diagram with
both a SG/PM second order and a first order phase transition (FOPT),
i.e., displaying latent heat and phase coexistence
\cite{footnote3}. The latter is characterized by the phenomenon of IT
\cite{CLPRL05,LPM}.
This is at contrast with the behavior of the original ordered BC
models, in which no IT was observed \cite{CapelPhys66,BEGPRA71,Saul74}
and with the behavior of a 3D BC model with quenched disorder on a
hierarchical lattice \cite{Ozcelik08}, yielding no IF, nor first order
SG/PM transition.
Eventually, to the best of our knowledge, the only claim of the
existence of a FOPT in $D=3$ systems in presence of quenched disorder,
has been made for the 4-Potts glass \cite{Fernandez08}.  In that case,
though, randomness tends to strongly smoothen the transition into a
second order one.
\\ \indent Motivated by the above considerations we have, thus,
numerically studied the existence
of IF  in a 3D random BC model with nearest-neighbor (nn)
interactions.
\\
\indent
{\em The  model.}
We consider the following Hamiltonian 
\BEQ
\mathcal{H}_J[s]=-\sum_{({i}{j})}
J_{{i}{j}}s_{{i}}s_{{j}}
+D\sum_{{i}}s^2_{{i}} \EEQ
where $({i}{j})$ indicate ordered couples of nn sites, and $s_{{i}}=
-1,0,+1$ are spin-1 variables lying on a cubic lattice of size $N=L^3$
with periodic boundary condition.  Crystal field $D$ is a chemical
potential for the magnetically active sites. Random couplings
$J_{{i}{j}}$ are independent identically distributed as $
P(J_{{i}{j}})=1/2~ \delta (J_{{i}{j}}-1) +1/2~\delta(J_{{i}{j}}+1)$.
%
 We simulate two real replicas $\{s^{(1)}_i\}$ and $\{s_i^{(2)}\}$ of
the system and define the overlap,
order
parameter of SG transition, as
$ q^{(J)}\equiv
1/N\sum_{{i}}\langle s^{(1)}_{{i}}
s^{(2)}_{{i}}\rangle$,
 where $\langle \ldots \rangle$ is the thermal average. 
If a FOPT occurs, the order parameter
  characterizing the transition is the density  of magnetically
  active ($|s_i|=1$) sites:
$\rho^{(J)}=1/N\sum_{{i}} \langle s^2_{{i}}\rangle$
or, since we deal with finite size (FS) systems, its distribution
$P_{N,J}(\rho)$. The values of the parameters depend on the particular
realization of disorder ($\{J_{{i}{j}}\}$). Such dependence is
self-averaging for the density probability distribution ($P_{N,J}(\rho)\sim 
P_N(\rho)\equiv {\overline {P_{N,J}(\rho)}}$ for $N\gg 1$),
 but not for the overlap distribution. We
denote by ${\overline {\phantom{(}\ldots \phantom{)}}}$ the average
over quenched disorder.
\\ \indent {\em Finite Size Scaling (FSS) for Continuous Transitions.}~ In
order to infer the details of the critical behavior from numerical
simulations of FS systems, a fundamental quantity is the four-point
correlation function, i.e., the correlation between local overlaps
$q_i = s^{(1)}_{{i}} s^{(2)}_{{i}}$:
\BEA
C_4({r})&\equiv&\frac{1}{N}\sum_{{i}}\overline{\langle
q_{{i}}~q_{{i}+{r}}\rangle} 
\EEA
The information contained in $C_4$ can be exploited to identify the
 existence of a second order phase transition for FS systems, e.g.,
 looking at a FS correlation length-like scaling function
 defined, on a 3D lattice, as \cite{Caracciolo93}:
\BEQ
\xi_c^2=\frac{1}{4\sin^2{{k_1/2}}}
\left(\frac{\hat{C}_4(0)}{\hat{C}_4({k_1})}-1\right)
\label{f:xi}
\EEQ
 where $\hat{C}_4(k)$ is the Fourier transform of $C_4({r})$,
$k_1=|\underline{k}_1|$, and $\underline{k}_1\equiv(\frac{2\pi}{L},0,0)$
is the minimum wave-vector.  In the thermodynamic
limit, a second order transition is characterized by a diverging
correlation lenght, at critical temperature $T_c$, whose FSS behavior
is the same as in Eq. (\ref{f:xi}) \cite{palassini,BCPRB00}.  Another
relevant observable is the SG susceptibility $\chi_{SG}\equiv N
\overline{\langle q^2 \rangle}=\hat{C}_4(0)$, diverging at the PM/SG
transition as $N \to\infty$. Because of FS, though, $\xi_c$ and
$\chi_{SG}$ cannot diverge in numerical simulations.  Around the
critical region, however, scale invariance survives. In fact, we can
define a FS ``critical'' temperature $T_c^L$ as the temperature at
which the above mentioned observables do not depend on the size.  In
this scaling region we have
\BEA
\xi_c/L&=&\bar \xi_c(\xi_c/L)= \bar\xi(L^{1/\nu}(T-T_c))
\\
\chi_{SG} L^{\eta-2}&=&\bar \chi(\xi_c/L) = \bar \chi(L^{1/\nu}(T-T_c))
\EEA
The critical temperature can, then, be estimated by FSS
of $T_c^L$ in the $L\to \infty$ limit.
\\ \indent In order to estimate the critical exponents we use the FSS
quotient method \cite{BCPRB00}, based on the observation that at
$T_c^L$ the correlation lenghts of different linear sizes
$L$ and $sL$ (in $L$ units) are equal:
$s\,\xi_c(T_c^L,L) = \xi_c(T_c^L,sL)$. 
%
For an observable $A$ diverging as $t^{x_A}$ ($t=T/T_c-1$) this implies:
 \BEQ
s^{\frac{x_A}{\nu}}=\frac{A(T_c^L,sL)}{A(T_c^L,L)}+\mathcal{O}(L^{-\omega})
\label{eq:quotient}.\EEQ 
 For a SG we can obtain the
exponents $\nu,\,\eta$ by means  of the FSS of the quotients of
$\p_\beta\xi_c$ and $\chi_{SG}$, scaling, respectively, with exponents
$x_{\p_\beta\xi}=1+\nu$ and 
$x_{\chi_{SG}}=(2-\eta)\nu $.
\\
\indent
{\em Characterization of  First Order Transitions.}~
The Clausius-Clapeyron equation for our model, where $D$ plays the
role of a pressure, reads~\cite{CLPRL05}
\BEQ
\frac{dD}{dT}=\frac{s_{PM}-s_{SG}}{\rho_{PM}-\rho_{SG}}=\frac{\Delta
s}{\Delta\rho}.
\label{eq:CC}
 \EEQ 
When the system undergoes a FOPT, a discontinuous
jump in $\rho$ (and, thus, in $q$) occurs. At finite $N$, $P_{N}(\rho)$
displays two peaks  in the coexistence
region corresponding  to PM ($\rho_{\rm PM}$) and
 SG ($\rho_{\rm SG}$)
phases.
 The FS transition line $D_c(N,T)$ can be evaluated as the locus of
points where the two phases are equiprobable, i.e., the areas of the
two peaks are equal \cite{Hill}:
\BEQ 
\int_{0}^{\rho_0} d\rho\,P_N(\rho)=\int_{\rho_0}^1
 d\rho\, P_N(\rho), 
\label{eq:equalarea} \EEQ 
with $\rho_0\in [\rho_{PM},\rho_{SG}]$
 such that $P_N(\rho_0)=0$ (or minimal for small $N$ 
next to the tricritical point).
\\
\indent 
{\em Exchange Monte Carlo  in T and D.}~
We simulated the equilibrium dynamics of our model
 using the parallel tempering (PT) algorithm, replicating
 the system 
both on different parallel
 $T$ and  $D$.
For the PT in $T$, the swap probability of two
copies between  $T=T$ and $T+\Delta T$ is:
$P_{swap}(\Delta \beta)=
\min\left[1,\exp\{\Delta\beta\Delta\mathcal{H}\}\right] $.
Between $D$ and $D+\Delta D$ it  reads 
$P_{swap}(\Delta D)=\min\left[1,\exp\{\beta\Delta
D\Delta\rho\}\right]$.
We used the latter to identify the reentrance of the
transition line in the $T,D$ phase diagram, cf. Fig. \ref{fig:phdi}.
\\ \indent We studied 3D systems with PT in $T$ at $D=0,1,1.75,2,2.05,
2.11$, and in $D$ at $T=0.2,0.3,0.4,0.5$. At all $D$ we simulated from
$33$ to $40$ replicated thermal baths $N_T$ at linear size
$L=6,8,10,12$ (number of disordered sample: $N_J=2000$). For
$D=0,1,1.75,2$ we simulated $N_T\in [20:33]$ at $L=16,20$
($N_J\in[900:1500]$) and $N_T\in [17:22]$ at $L=24$
($N_J\in[500:1000]$).  For the PT cycles in $D$, $N_D\in[21:37]$,
parallel replicas at different $D$ were simulated, of 
$L=6,8,10,12$ and $15$ ($N_J=1000$). In the latter case varying
$\Delta D$ were used, larger in the pure phases and progressively
smaller approaching the coexistence region.  The number of
Monte Carlo (MC) steps varies  from $2^{15}$ to $2^{21}$ according to $L$
and to the lowest
values of $T,D$ reached.
Thermalization has been checked by looking at: (i) the symmetry of the
overlap distributions $P_{N,J}(q)$, (ii) the $t$-log behavior of the
energy (when at least the last two points coincide), (iii) the lack of
variation of each considered observable (e.g., $\xi, \chi_{SG}$) on
logarithmic time-windows.
\\ \indent {\em Numerical results.}~ 
In Fig.  \ref{fig:xisuL} we present the $T$-behavior of $\xi/L$ for
$D=0,1,2,2.11$.  From the FSS analysis of their crossing points we can
determine the critical temperature and, applying the quotient method
(cf. Eq. (\ref{eq:quotient}) with $s=2$), we obtain estimates for the
critical exponents: $\nu = 2.44(6) $, $\eta = -0.34(2)$ at $D=0$, $\nu
= 2.4(2)$, $\eta = -0.31(2)$ at $D=1$, $\nu = 2.1(2)$, $\eta =
-0.27(2)$ at $D=1.75$.  The system appears to be in the same
universality class of the Edwards-Anderson model (corresponding to the
$D=-\infty$ limit of our model) \cite{BCPRB00,Joerg06}. For $D=2$,
near the tricritical point ($0.5\lesssim T \lesssim 0.54, 2.05
\lesssim D \lesssim 2.11$), the quotient method does not yield
reliable estimates because of a crossover in the scaling functions in
the range of probed sizes ($L=6$ to $24$). This comes about because at
the tricritical point the coefficient of the fourth order term in the
SG free energy action goes to zero and the sixth order term becomes
relevant for the critical behavior \cite{CLPRL02}, a typical behavior
of BEGC-like systems \cite{Riedel72}.
\\ \indent At $D= 2.11$ no evidence is found for a second order phase
transition whereas a FOPT is observed at $T=0.51(1)$.  This is also
the reason why, as shown in the bottom-right panel of
Fig. \ref{fig:xisuL}, $\xi_c/L$ decays in the coexistence region, for
$T\lesssim 0.51$.
\begin{figure}[t!] 
\centering
\includegraphics[width=.9\columnwidth]{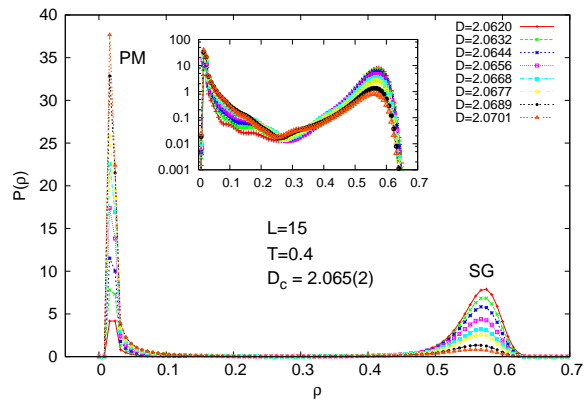}
\caption{Density distribution $P_L(\rho)$, $L=15$, across the
coexistence region at $T=0.4$: two peaks develop at $\rho_{PM}$ and
$\rho_{SG}$. As $D$ increases the thermodynamically relevant phase
(lowest free energy) passes from SG to PM in a first order phase
transition. The dominant phase correpsonds to the one with larger
probability, i.e., larger intergral of the peak. As the peak at
$\rho_{SG}$ vanishes the system is in a purely PM phase.
Inset: $P_{15}(\rho)$ on $y$-Log scale.}
\label{fig:Prho}
\end{figure}
\\ \indent The FOPT is determined by looking at $P_L(\rho)$ as two
peaks appear, and computing the $(D,T)$ points satisfying
Eq. (\ref{eq:equalarea}).  A pure phase corresponds to a single-peaked
distribution, whose peak is narrower for larger systems.  In
Fig. \ref{fig:Prho} we show the behavior of the $P_{15}(\rho)$ through
the FOPT in $D$ at $T=0.4$ \cite{footnote5}.
  A detail of the FSS limit of the IT line $D_c(T)$
and of the spinodal lines is plotted in the bottom inset of
Fig. \ref{fig:phdi}.  The spinodal lines at given $L$ are estimated by
looking at the $D$ values at which a secondary peak first arises.
 \\ \indent Using Eq. (\ref{eq:CC}), from the knowledge of $\Delta
\rho$ and the estimate of $dD/dT$ by numerical interpolation we
compute the latent heat $|\Delta s|/T =(s_{SG}-s_{PM})/T$ employed in
the transition, cf.  bottom inset of Fig. \ref{fig:phdi}: as $T$ {\em
increases} the PM acquires latent heat to {\em vitrify}; SG entropy is
higher than $s_{PM}$ and the "frozen" phase is found at a higher $T$
than the fluid one.
The IF takes place between a SG of high density to an almost empty PM 
(e.g., at T=0.4, in the coexistence region
$D\in[2.046(2):2.092(5)]$, $\rho_{SG}\simeq 0.52$ and $\rho_{PM}\simeq
0.03$). The few active sites do not interact with each other but only
with inactive neighbors and this induces zero magnetization and
overlap. The corresponding PM phase at high $T$ has, instead, higher
density (e.g., $\rho_{PM}(D=2,T=0.6)=0.4157(2)$,
$\rho_{PM}(D=2.11,T=0.6)=0.596(2)$) and the paramagnetic behavior is
brought about by the lack of both magnetic order (zero magnetization)
and blocked spin configurations (zero overlap).
\\ \indent {\em Conclusions.}~We focused on a
spin-1 SG model on a 3D cubic lattice, whose constituent features try to
capture at least one supposed mechanism underlying IT: the raise of
inactive components at low $T$. We provide numerical evidence
for an equilibrium {\em inverse freezing} phenomenon: at given values
of an external field, heating up a paramagnet this is transformed into
a SG.  The whole phase diagram in temperature $T$ and crystal field
$D$  has been studied both along the continuous transition
line, where critical exponents are computed, and in the coexistence
region, where FOPT line, latent heat curve and
spinodal lines are reported. This latter observations confirm the claim
of Fernandez {\em et al.} \cite{Fernandez08} about the existence of
such transitions in quenched disordered short-range finite-dimensional
systems.  In the present model the FOPT can be seen by
means of standard PT in the canonical ensemble,
simply tuning an external pressure-like parameter (that is, a
parameter adjustable in experiments on real samples \cite{lastfootnote}).  Besides the
peculiarity of FOPT, inverse freezing is also observed, for the first
time, in a random short-range finite dimension system.
 Both features were absent in the same model
on a hierarchical lattice \cite{Ozcelik08} and in the ordered BC model no IT 
was observed 
\cite{CapelPhys66,Saul74}. 
\\
 \indent
 {\em Acknowledgments.}~ Funding for this work has been provided by
the INFM-CNR Seed Grant ``Order in disorder''.  We thank
A. Nihat-Berker and H. Katzgraber for useful discussions.


\begin{thebibliography}{99}
\bibitem{Wilks87} J. Wilks and D.S. Betts, {\em An introduction to liquid Helium}, Clarendon Press (Oxford, 1987).
\bibitem{Chevillard97} C. Chevillard and M.A.V. Axelos,
Colloid. Pol. Sci. {\bf 275}, 537 (1997).

\bibitem{RHKMM99} S. Rastogi, G.W.H. H{\"o}hne and A. Keller,
Macromolecules {\bf 32}, 8897 (1999); A.L. Greer, Nature {\bf 404},
134 (2000); N.J.L. van Ruth and S. Rastogi, Macromolecules {\bf 37},
8191 (2004).

\bibitem{Plazanet}
M. Plazanet {\em et al.}
J. Chem. Phys. {\bf 125} 154504 (2006).
R. Angelini and G. Ruocco, Phil. Mag. B {\bf 87},
553 (2007).  R. Angelini, G. Salvi and G. Ruocco, Phil. Mag. B {\bf
88}, 4109 (2008).  R. Angelini, G. Ruocco, S. De Panfilis,
Phys. Rev. E {\bf 78}, 020502(R) (2008).
E. Tombari {\em et al.}, J. Chem. Phys. {\bf 123},
051104 (2005).
 
\bibitem{Donnio07} B. Donnio {\em et al.},
Adv. Mater. {\bf 19}, 3534 (2007).

\bibitem{Avraham01}    N. Avraham {\em et al.},  Nature  {\bf 411} 451 (2001).

\bibitem{Tammann} G. Tammann, ``Kristallisieren und Schmelzen'',
Metzger und Wittig, Leipzig (1903).

\bibitem{footnote1} Generally speaking, at a fixed parameter {\em
 externally} tuning the interaction strength such as concentration,
 chemical potential or magnetic field.

\bibitem{Cladis75} P.E. Cladis, Phys. Rev. Lett. \textbf{35}, 48 (1975): \textbf{39}, 720 (1977).
 H. {\"O}zbek {\em et al.}, Ph. Trans. {\bf 75}, 301 (2002).
A. Srivastava, D. Sa and S. Singh, Eur. Phys. J. E {\bf 22}, 111 (2007) and references therein.
O. Portmann, A. Vaterlaus, and D. Pescia,  Nature {\bf 422}, 701 (2003).

\bibitem{Arisz95} P.W. Arisz, H.J.J. Kauw and J.J. Boon, Carbohydr. Res. \textbf{271}, 1 (1995).
\bibitem{Haque93} A. Haque and E.R. Morris, Carbohyd. Pol. {\bf 22},
161 (1993).
\bibitem{Sciortino09} F. Sciortino, A. Giacometti and G. Pastore,
Phys. Rev. Lett.  \textbf{103} 237801 (2009).
 \bibitem{SSPRL04}
  N. Schupper and N.M. Shnerb, Phys. Rev. Lett.  {\bf 93} (2004)
  037202.  
\bibitem{SS05} N. Schupper and N.M. Shnerb,  Phys. Rev. E {\bf 72},  046107 (2005).
\bibitem{Feeney}
M.R. Feeney., P.G. Debenedetti, and F.H. Stillinger,
 J. Chem. Phys. {\bf 119} 4582 (2003).
S. Prestipino, Phys. Rev. E {\bf 75}, 011107 (2007).

\bibitem{CLPRL05}  A. Crisanti and L. Leuzzi, Phys. Rev. Lett. {\bf 95},
 087201 (2005).
\bibitem{Sellitto06} M. Sellitto,  Phys. Rev. B {\bf 73} 180202(R) (2006).

\bibitem{GSJPC77} S. K. Ghatak, D. Sherrington, J. Phys. C: Solid State Phys.
{\bf{10}}, 3149 (1977).

\bibitem{CLPRL02} A. Crisanti and  L. Leuzzi,
Phys. Rev. Lett. {\bf 89} (2002) 237204;  Phys. Rev. B {\bf 70}
(2004) 014409.
\bibitem{footnote3} The transition is first order in the {\em
thermodynamic} sense and is not related to the so-called random first
order transition occurring in mean-field models for structral
glasses.
\bibitem{LPM} L. Leuzzi, \emph{Phil. Mag.} \textbf{87}, 543-551
(2006).


\bibitem{BEGPRA71} M. Blume, V.J. Emery and R.B. Griffiths,
 Phys. Rev. A {\bf{4}},  1071 (1971).

\bibitem{CapelPhys66}H. W. Capel, Physica {\bf{32}}, 966 (1966);
M. Blume, Phys. Rev.  {\bf{141}}, 517 (1966).


\bibitem{Saul74} D.M. Saul, M. Wortis and D. Stauffer, Phys. Rev. B {\bf  9}, 4964 (1974).
A. Nihat Berker and M. Wortis, Phys. Rev. B {\bf 14}, 4946 (1976).
 A. K. Jain and D. P. Landau, Phys. Rev. B {\bf 22}, 445 (1980).
 


\bibitem{Ozcelik08} V.O. {\"O}z\c{c}elik and A. N. Berker,
Phys. Rev. E {\bf 78}, 031104 (2008).

\bibitem{Fernandez08} L.A. Fern{\`a}ndez {\em et al.},
Phys. Rev. Lett. {\bf 100}, 057201 (2008).


\bibitem{Caracciolo93}S. Caracciolo {\em et al.}, Nucl. Phys. B403, 475
  (1993).
\bibitem{BCPRB00} H. G. Ballesteros {\em et al.},
Phys. Rev. B {\bf 62} (2000) 14237.
\bibitem{palassini} M. Palassini, S. Caracciolo,
Phys. Rev. Lett. \textbf{82}, 5128 (1999).

\bibitem{Hill} T.L. Hill, {\em Thermodynamics of Small Systems}, Dover (2002).


\bibitem{Joerg06} T. J\"org, Phys. Rev. B \textbf{73}, 224431
  (2006). H.G. Katzgraber, M. Korner and A.P. Young, Phys. Rev. B
  {\bf 73}, 224432 (2006). M. Hasenbusch {\em et al.},
 Phys. Rev. B  \textbf{76}, 094402 (2007); {\em ibid.} \textbf{76} 184202 (2007).

\bibitem{Riedel72} E.K. Riedel and F.J. Wegner, Phys. Rev. Lett. \textbf{29}, 349 (1972). J. Zinn-Justin, {\em Quantum Field Theory and Critical Phenomena}, Oxford University Press (Oxford, 1989).


\bibitem{footnote5} Though $P_L(\rho)$ is not
exactly zero between the two peaks, cf. inset of
Fig. \ref{fig:Prho}, the determination of $D_c(T)$ with the present precision is robust against changes of $\rho_0$ in a reasonably wide domain of $\rho$.


\bibitem{lastfootnote}
Although in numerical simulations
         changing the pressure or, e.g., the bond dilution  \cite{Fernandez08}, or the
         relative probabilities of the random bond values \cite{Toldin09} 
         is technically equivalent, the latter two cannot be tuned in a single real thermodynamic experiment and require the
         preparation of several samples with different microscopic
         properties.  
         
\bibitem{Toldin09}  F.P. Toldin, A. Pelissetto and E. Vicari, J.  Stat. Phys. {\bf 135},
         1039 (2009).

 \end{thebibliography}
\end{document}